\documentclass[fleqn,twoside]{article}
\usepackage{espcrc2,epsfig}
\usepackage{graphicx}
\usepackage[figuresright]{rotating}

\newcommand{\AmS}{{\protect\the\textfont2
  A\kern-.1667em\lower.5ex\hbox{M}\kern-.125emS}}

\textfloatsep=\floatsep           % Space between main text and floats
\intextsep=\floatsep

\title{Present status of IGEX dark matter search at Canfranc Underground Laboratory}

\author{I.G. Irastorza\address[UZ]{\footnotesize Laboratory of Nuclear and High Energy Physics, University of
Zaragoza, 50009 Zaragoza, Spain}\thanks{Attending speaker:
Igor.Irastorza@cern.ch}\thanks{Present address: CERN, EP Division,
CH-1211 Geneva 23, Switzerland},A. Morales\addressmark, C.E.
Aalseth\address[USC]{\footnotesize University of South Carolina,
Columbia, South Carolina 29208 USA}, F.T. Avignone
III\addressmark, R.L. Brodzinski\address[PNL]{\footnotesize
Pacific Northwest National Laboratory, Richland, Washington 99352
USA}, S. Cebri\'{a}n\addressmark[UZ], E.~Garc\'{\i}a\addressmark[UZ],
I.V. Kirpichnikov\address[ITEP]{\footnotesize Institute for
Theoretical and Experimental Physics, 117 259 Moscow, Russia},
A.A. Klimenko\addressmark, H.S. Miley\addressmark[PNL], J.
Morales\addressmark[UZ], A. Ortiz de Sol\'{o}rzano\addressmark[UZ],
S.B.~Osetrov\address[INR]{\footnotesize Institute for Nuclear
Research, Baksan Neutrino Observatory, 361 609 Neutrino, Russia},
V.S. Pogosov\address[YPI]{\footnotesize Yerevan Physical
Institute, 375 036 Yerevan, Armenia}, J. Puimed\'{o}n\addressmark[UZ],
J.H.\ Reeves\addressmark[PNL], M.L. Sarsa\addressmark[UZ],
A.A. Smolnikov\addressmark[INR], A.G.~Tamanyan\addressmark[YPI],
A.A. Vasenko\addressmark[INR], S.I. Vasiliev\addressmark[INR],
J.A. Villar\addressmark[UZ].}

\begin{document}

\begin{abstract}
One IGEX $^{76}$Ge double-beta decay detector is currently
operating in the Canfranc Underground Laboratory in a search for
dark matter WIMPs, through the Ge nuclear recoil produced by the
WIMP elastic scattering. A new exclusion plot, $\sigma$(m), has
been derived for WIMP-nucleon spin-independent interactions. To
obtain this result, 40 days of data from the IGEX detector (energy
threshold $E_{thr} \sim 4$ keV), recently collected, have been
analyzed. These data improve the exclusion limits derived from all
the other ionization germanium detectors in the mass region from
20~GeV to 200~GeV, where a WIMP supposedly responsible for the
annual modulation effect reported by the DAMA experiment would be
located. The new IGEX exclusion contour enters, by the first time,
the DAMA region by using only raw data, with no background
discrimination, and excludes its upper left part. It is also shown
that with a moderate improvement of the detector performances, the
DAMA region could be fully explored.
\end{abstract}

\maketitle

\section{Introduction}
%Substantial evidence exists suggesting most matter in
%the universe is dark, and there are compelling reasons to believe
%it consists mainly of non-baryonic particles. Among these candidates,
%Weakly Interacting
%Massive and neutral Particles (WIMPs) are among the
%front runners. The lightest stable
%particles of supersymmetric theories, like the neutralino, describe a
%particular
%class of WIMPs \cite{Gri}.

Recent cosmological observations and robust theoretical arguments
require an important Dark Matter component ($\Omega_{DM}\sim
25-30\%$) in our universe, which is supposed to be made mostly of
non-baryonic particles. Weakly Interacting Massive (and neutral)
Particles (WIMPs), which are favourite candidates to such
non-baryonic component, could fill the galactic halos accounting
for the flat rotation curves which are measured for many galaxies.
They could be detected by measuring the nuclear recoil produced by
their elastic scattering off target nuclei in a suitable detector
\cite{Mor99}. In this talk new WIMP constraints derived from a
germanium detector of the IGEX collaboration are presented. They
improve previous limits obtained with Ge ionization detectors, and
enter by the first time the so-called DAMA region (corresponding
to a WIMP supposedly responsible for the annual modulation effect
found in the DAMA experiment \cite{Ber99}) without using
mechanisms of background rejection, but relying only on the
ultra-low background achieved.

\section{Experiment}
The IGEX experiment \cite{Aal,Gon99}, optimized for detecting
$^{76}$Ge double-beta decay, has been described in detail
elsewhere. One of the IGEX detectors of 2.2 kg (active mass $\sim$
2.0 kg), enriched up to 86 \% in $^{76}$Ge, is being used to look
for WIMPs interacting coherently with the germanium nuclei. The Ge
detector and its cryostat were fabricated following
state-of-the-art ultralow background techniques and using only
selected radiopure material components (see Ref \cite{Aal,Mor00}).

The detector shielding has been recently modified with respect to
that of the previous set-up of Ref. \cite{Mor00}. The improvements
concern basically the neutron moderator. We have doubled its
thickness (40 cms of polyethylene and borated water tanks) and now
it much better covers the whole set-up, due to the removal of the
other detectors (RG-I, RG-III and COSME), which dewars did not
allow us to perfectly close the polyethylene wall. These changes
were motivated by a previous study of the possible sources of
background in IGEX based on several simulations, which pointed out
that the neutrons from the surrounding rock could contribute
considerably to the low energy background. For more details on the
IGEX shielding we refer to \cite{Mor00} and \cite{IGEX2001}.

In addition to the data acquisition system used in previous runs
(described in \cite{Mor00}), a specific pulse shape analysis has
been implemented for the data set presented in this talk. The
pulse shapes of each event before and after amplification are
recorded by two 800 MHz LeCroy 9362 digital scopes. These are
analyzed one by one by means of a method based on wavelet
techniques which allows us to assess the probability of this pulse
to have been produced by a random fluctuation of the baseline.
This probability is used as a criterium to reject events coming
from electronic noise or microphonics. According to the
calibration of the method, it works very efficiently for events
above 4 keV.

\begin{figure}[t]
\centerline{ \epsfxsize=7.5cm \epsffile{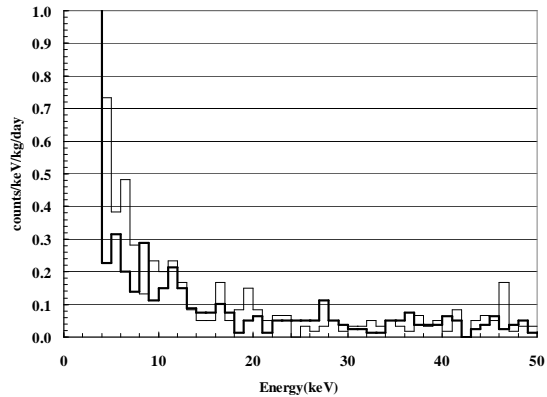} }
 \caption{\footnotesize Normalized low energy spectrum of the IGEX RG-II detector
 corresponding to the 80~kg~d presented in this talk (thick line)
 compared to the previous 60~kg~d spectrum published in \cite{Mor00}.}
 \label{dm-ig-1}
\end{figure}

\begin{figure}[t]
\centerline{ \epsfxsize=7.5cm \epsffile{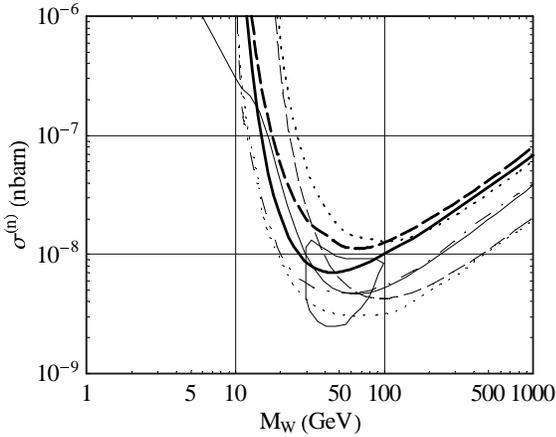} }
 \caption{\footnotesize IGEX-DM exclusion plot for spin-independent
interaction obtained in this work (thick solid line) compared with
the previous exclusion obtained by IGEX-DM \protect\cite{Mor00}
(dashed thick line) and the last result obtained by the
Heidelberg-Moscow germanium experiment \cite{Bau} (dotted line)
recalculated from the original spectrum with the same set of
hypothesis and parameters. The closed line corresponds to the
(3$\sigma$) annual modulation effect reported by the DAMA
collaboration (including NaI-1,2,3,4 runnings) \cite{Ber99}. The
thin solid line is the exclusion line obtained by DAMA NaI-0
\cite{Ber96} by using Pulse Shape Discrimination. The two other
experiments which have entered the DAMA region are also shown:
EDELWEISS \cite{Benoit:2001} (thin dashed line) and the CDMS
%exclusion contour (thin dotted line) \cite{Abusaidi:2000} and its
%expected sensitivity \cite{Sadoulet:2001} (thin dot-dashed
%line)
(For which two contour lines habe been depicted according to a
recent recommendation \cite{Sadoulet:2001}, the exclusion plot
published in Ref. \cite{Abusaidi:2000} --thin dotted line-- and
the CDMS expected sensitivity contour \cite{Abusaidi:2000} --thin
dot-dashed line--).
%Also
%shown are the other three experiments which enters in the DAMA
%region: CDMS \cite{Abusaidi:2000} (dot-dashed thin line),
%EDELWEISS \cite{Benoit:2001} (dashed thin line) and DAMA NaI-0
%\cite{Ber96} (thin solid line).
} \label{dm-ig-2}
\end{figure}

\begin{figure}[t]
\centerline{ \epsfxsize=7.5cm \epsffile{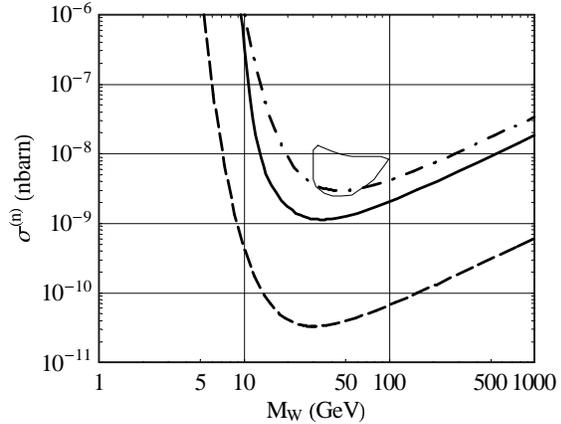} }
 \caption{\footnotesize IGEX-DM projections are shown for
 a flat background rate of 0.1~c/keV/kg/day (dot-dashed line) and 0.04~c/keV/kg/day (solid line) down
 to the threshold at 4 keV, for 1~kg~year of exposure.
 The exclusion contour expected for GEDEON is also
 shown (dashed line) as explained in the text.} \label{prospects}
\end{figure}

\section{Results and prospects}

The results presented in this talk are from a recent run of an
exposure of Mt=80 kg~days with the modified set-up previously
mentioned. The spectrum obtained is shown in Figure~\ref{dm-ig-1}
compared with the previous IGEX published spectrum of Ref.
\cite{Mor00}.
%Also presented for comparison
%are recent COSME detector results obtained in its current set-up
%(COSME-2) \cite{Ceb00,Mor99}. Earlier results from the COSME
%detector (COSME-1) \cite{Jmor,Gar92} are included in the
%Ge-combined bound contributing mainly to the low-mass WIMP region
%of the exclusion plot because of its low energy threshold.

The energy threshold of the detector is 4 keV and the FWHM energy
resolution at the 75 keV Pb X-ray line was of 800 eV. The
background rate recorded was $\sim$0.21, $\sim$0.10 and $\sim$0.04
c/keV/kg/day between 4--10~keV, 10--20~keV, and 25--40~keV
respectively. As it can be seen, the background below 10 keV has
been substantially reduced (about a factor 50\%) with respect to
that obtained in the previous set-up \cite{Mor00}, essentially due
to the improved shielding (both in lead and in
polyethylene-water).
%It is worth mentioning
%that this reduction was not due to the implementation of the Pulse
%Shape Analysis, which left, in fact, the spectrum shape roughly
%unchanged.
This suggests that the neutrons could be an important component of
the low energy background in IGEX.

The exclusion plot is derived from the recorded spectrum following
the same set of hypothesis and parameters used in previous papers
(see \cite{Mor00}) and is shown in Fig.~\ref{dm-ig-2} (thick solid
line). It improves the IGEX-DM previous result (thick dashed line)
as well as that of the other previous germanium ionization
experiments (including the last result of Heidelberg-Moscow
experiment \cite{Bau} --thick dotted line--) for a mass range from
20~GeV to 200~GeV, which encompass that of the DAMA mass region.
%To compare the IGEX exlcusion plots with that derived from the
%Heidelberg/Moscow data \cite{Bau}, the recoil energy dependent
%ionization yield used is the same that in Ref. \cite{Bau}, E$_{\rm
%vis}=0.14\ (\rm E_{recoil})^{1.19}$.
In particular, this new IGEX result excludes WIMP-nucleon
cross-sections above 7$\times 10^{-9}$ nb for masses of $\sim$50
GeV and enters the so-called DAMA region \cite{Ber99} where the
DAMA experiment assigns a WIMP candidate to their found annual
modulation signal. IGEX excludes the upper left part of this
region. That is the first time that a direct search experiment
without background discrimination, but with very low (raw)
background, enters such region.
%Also shown is the exclusion contours derived from the last
%Heidelberg-Moscow result \cite{Bau} and that obtained from the
%previous IGEX result. All have been recalculated from the original
%energy spectra with the same set of hypotheses and parameters.
Also shown for comparison are the contour lines of the other
experiments which have entered that region.

%depicted according to a recent recommendation
%\cite{Sadoulet:2001}, the exclusion plot published in Ref.
%\cite{Abusaidi:2000} (thin dotted line) and the CDMS expected
%sensitivity contour \cite{Abusaidi:2000} (thin dot-dashed line)}
%\cite{Abusaidi:2000} and EDELWEISS \cite{Benoit:2001} (thin dashed
%line), which have entered that region, and the exclusion plot
%obtained by DAMA NaI-0 (thin solid line) \cite{Ber96} by using
%statistical pulse shape discrimination.

% The DAMA experiment contour
%line (thin solid line) derived from Pulse Shape Discriminated
%spectra \cite{Ber96}, and the DAMA (``triangle'') region
%corresponding to its reported annual modulation effect
%\cite{Ber99} are also shown.

%The IGEX exclusion contour improves on that of other germanium
%experiments for masses corresponding to that of the neutralino
%tentatively assigned to the DAMA modulation effect \cite{Ber99}
%and results from using raw data without background subtraction. It
%also excludes the upper left part (see Fig. \ref{dm-ig-2}) of the
%DAMA region. The plots obtained by using instead other
%energy-dependent expressions for the ionization yield, like that
%given by Smith and Lewin \cite{Smith} do not change appreciably
%the exclusion plots of Fig. \ref{dm-ig-2}.

Data collection is currently in progress and some strategies are
being considered to further reduce the low energy background. If
this reduction is achieved, very interesting perspectives can be
set for IGEX. In Fig.~\ref{prospects} we plot the exclusions
obtained with a flat background of 0.1 c/kg/keV/day (dot-dashed
line) and of 0.04 c/kg/keV/day (solid line) down to the current 4
keV threshold for an exposure of 1~kg~year. In particular, the
complete DAMA region could be tested with a moderate improvement
of the IGEX performances. The dashed line in Fig.~\ref{prospects}
corresponds to a flat background of 0.002 c/kg/keV/day down to a
threshold of 4 keV and 24 kg y of exposure, which are the
parameters expected for GEDEON (GErmanium DEtectors in ONe
cryostat), a new experimental project on WIMP detection using
larger masses of natural germanium planned as an extension of the
IGEX dark matter search (see ref.~\cite{IGEX2001}). GEDEON would
be massive enough \cite{Cebrian:2001} to search also for the WIMP
annual modulation effect and explore positively an important part
of the WIMP parameter space including the DAMA region.

%
%\section*{Acknowledgements}
%The Canfranc Astroparticle Underground Laboratory is operated by
%the University of Zaragoza under contract No. AEN99-1033. This
%research was funded by the Spanish Commission for Science and
%Technology (CICYT), the U.S. National Science Foundation, and the
%U.S. Department of Energy. The isotopically enriched $^{76}$Ge was
%supplied by the Institute for Nuclear Research (INR), Moscow, and
%the Institute for Theoretical and Experimental Physics (ITEP),
%Moscow.

\end{document}